\documentclass{jfm}
\usepackage{graphicx}
\usepackage{natbib}

\ifCUPmtlplainloaded \else
  \checkfont{eurm10}
  \iffontfound
    \IfFileExists{upmath.sty}
      {\typeout{^^JFound AMS Euler Roman fonts on the system,
                   using the 'upmath' package.^^J}%
       \usepackage{upmath}}
      {\typeout{^^JFound AMS Euler Roman fonts on the system, but you
                   dont seem to have the}%
       \typeout{'upmath' package installed. JFM.cls can take advantage
                 of these fonts,^^Jif you use 'upmath' package.^^J}%
      }
  \else
  \fi
\fi

\ifCUPmtlplainloaded \else
  \checkfont{msam10}
  \iffontfound
    \IfFileExists{amssymb.sty}
      {\typeout{^^JFound AMS Symbol fonts on the system, using the
                'amssymb' package.^^J}%
       \usepackage{amssymb}%
         \let\leq=\leqslant
         
      }{}
  \fi
\fi

\ifCUPmtlplainloaded \else
  \IfFileExists{amsbsy.sty}
    {\typeout{^^JFound the 'amsbsy' package on the system, using it.^^J}%
     \usepackage{amsbsy}}
    {}
\fi

\newsavebox{\astrutbox}
\sbox{\astrutbox}{\rule[-5pt]{0pt}{20pt}}

\title[Axisymmetrically TC-like Vortices]{Axisymmetrically Tropical Cyclone-like Vortices with Secondary Circulations}

\author[L. Sun]%
{L\ls I\ls N\ls G\ns  S\ls U\ls N$^1$%
  \thanks{Corresponding Author: Liang Sun, email: sunl@ustc.edu.cn;\break sunl@ustc.edu.}
}

\affiliation{
$^1$School of Earth and Space Sciences, University of Science and
Technology of China, Jinzhai Road 96\#, Hefei, 230026, China.\\[\affilskip]
}

\date{\today}

\date{\today, and in revised form ??}

\begin{document}

\maketitle

\begin{abstract}
The secondary circulation of the tropical cyclone (TC) is related
to its formation and intensification, thus becomes very important
in the studies. The analytical solutions have both the primary and
secondary circulation in a three-dimensionally nonhydrostatic and
adiabatic model. We prove that there are three intrinsic radiuses
for the axisymmetrically ideal incompressible flow. The first one
is the radius of maximum primary circular velocity $r_m$. The
second one is radius of the primary kernel $r_k>r_m$, across which
the vorticity of the primary circulation changes sign and the
vertical velocity changes direction. The last one is the radius of
the maximum primary vorticity $r_d$, at which the vertical flow of
the secondary circulation approaches its maximum, and across which
the radius velocity changes sign. The first TC-like vortex
solution has universal inflow or outflow. The relations between
the intrinsic length scales are $r_k=\sqrt{2}r_m$ and $r_d=2r_m$.
The second one is a multi-planar solution, periodically in
$z$-coordinate. Within each layer, the solution is a convection
vortex. The number of the secondary circulation might be one, two,
three, and even more. There are also three intrinsic radiuses
$r_m$, $r_k$ and $r_d$, but they have different values. It seems
that the relative stronger radius velocity could be easily found
near boundaries. The above solutions can be applied to study the
radial structure of the tornados, TCs and mesoscale eddies.

\end{abstract}

\section{Introduction}
Typhoons (hurricanes or tropical cyclones) are intense atmospheric
cyclonic vortices that genesis over warm tropical oceans with
their energy primarily from the release of water vapor latent
heat. As the tropical cyclones (hereafter TCs) could make great
impacts on both oceanic and terrestrial environments, there were
lots of investigations on all the relative issues. Among them, a
very fundamental problem is understanding the three-dimensional
structures and dynamics of the TC, which is useful for both
weather forecasters and researchers. In this problem, the thermal
dynamics (moisture convection) and the mechanic dynamics are
strongly coupled, which could hardly be solved theoretically.

According to the observations, the TC has a strong cyclonic
circulation (primary circulation), and a weak vertical circulation
(secondary circulation). The inner part of the storm becomes
nearly axisymmetric as the storm reaches maturity, and its
strongest winds surround a relatively calm eye, whose diameter is
typically in the range of 20 to 100 km \cite[]{ChanBook2010}.
\cite{Nolan2002} pointed out that the TC can be represented to
zeroth order as a vortex in gradient wind and hydrostatic balance,
with no secondary circulation. In most theoretical studies, the
ideal axisymmetric and steady-state model was used. The model is
based on the assumptions that the flow above the boundary layer is
inviscid and thermodynamically reversible, that hydrostatic and
gradient wind balance could apply \cite[]{Emanuel1986,Stern2009}.
As a first step in the prediction of the TC track, the TC is taken
as a barotropic vortex. In some cases, a simple two-dimensional
barotropic ¡°dry¡± model is used to simulate a TC-like concentric
rings vortex \cite[]{Mallen2005,Martinez2010,Moon2010}.
\cite{Mallen2005} compared the different approximation of the
primary circulation with the observed data. The primary
circulation far away from the inner core is like a Rankine with
skirt vortex.

\iffalse \cite{Stern2009} found no relationship between the slope
of the the radius of maximum winds (RMW) and intensity. It is
shown that the RMW is indeed closely approximated by an absolute
angular momentum $M$ surface for the majority of storms. \fi

Another important feature of TC is the secondary circulations. The
secondary circulation is coupled to the vertical motion induced by
the convection that drives the vortex. Inside the radius of
maximum winds, where the radial inflow erupts out of the boundary
layer, vertical velocities are comparable to those in the overhead
convection \cite[]{Nolan2002}. The secondary circulation has
related to TC formation and intensification in idealized flow
configurations \cite[]{Montgomery2006,Kieu2009,Montgomery2011qj}.
Due to the complexity of the problem, \cite{Nolan2002} can only
compute the solutions of three-dimensional linear perturbations on
a primary circulation to study the secondary circulation.
\cite{Kieu2009} studied the evolution of a Rankine vortex with
exponential growth in the core region. However, due to the neglect
of both the vertical momentum equation and the thermodynamic
equation, the exponential growth rate of the TC seems to fast to
accept \cite[]{Montgomery2010qj}.

Though lots of works dealing with this problem, the dynamics and
the structure of the secondary circulation is far to a fine
diagram. In the schematic diagram of model for a mature
steady-state hurricane \cite[]{Emanuel1986}, the secondary
circulation is divided into three regions along radius. The inner
eye is Region I, then Region II is the eyewall cloud, the outer to
far is Region III. The model assumes that the radius of maximum
tangential wind speed, $r_m$, is located at the outer edge of the
eyewall cloud, whereas the recent observations indicate it is
closer to the inner edge \cite[]{Montgomery2011qj}.

As an alternative way to avoid the above disadvantage, we consider
to solve a nonhydrostatic and adiabatic model following the
deviation by \cite{Batchelor1967,Frewer2007} and to apply
separation of variables by \cite{SunL2011taml,SunL2011arx}. In
this study, we use the nonlinear Euler equations following the
assumptions by \cite{Emanuel1986}. And we consider the
axisymmetric vortices in a rotation plane with the constant
Coriolis parameter $f$ ($f$-plane). In the inner part of the TC
(e.g. the radius less than 50 km), the gradient balance dominates
the primary circulation \cite[]{Willoughby1990Jas}. As a result, a
general exact spiral solution is presented
in~\S\,\ref{sec:modelresults}, some three-dimensional TC-like
vortex solution with secondary circulation solutions are given
in~\S\,\ref{sec:sepcialsol}. Discussion and conclusion are
respectively given in~\S\,\ref{sec:discussion} and
in~\S\,\ref{sec:conclusion}, respectively.

\section{General solution} \label{sec:modelresults}

Different from the previous studies, a nonhydrostatic model is
employed to study the secondary circulation. We consider the
steady solutions of the incompressible Euler equations for
axisymmetric flow in a rotation plane. The constant $f$ represents
the Coriolis parameter due to rotation. It is convenient to use a
cylindrical coordinate system $(r,\theta,z)$ with the velocity
components ($V_r,V_{\theta},V_z$), and all the velocity components
are the functions of $r$ and $z$ but $\theta$, due to the
axisymmetric. As $V_r=V_r(r,z)$, $V_{\theta}=V_{\theta}(r,z)$ and
$V_z=V_z(r,z)$, the governing equations, including
mass-conservation and momentum equations, are:
\begin{subeqnarray}
\frac{\partial (rV_r)}{\partial r} + \frac{\partial (rV_z)}{\partial z} &=&0\\
V_r\frac{\partial V_{\theta}}{\partial r} + V_z\frac{\partial V_{\theta}}{\partial z}+\frac{ V_r V_{\theta}}{r} +fV_r &=&0\\
V_r\frac{\partial V_{r}}{\partial r} + V_z\frac{\partial V_{r}}{\partial z}-\frac{ V_{\theta}^2}{r}-f V_{\theta} -\frac{1}{4}f^2r &=&-\frac{1}{\rho}\frac{\partial p}{\partial r}\\
V_r\frac{\partial V_{z}}{\partial r} + V_z\frac{\partial
V_{z}}{\partial z} &=&-\frac{1}{\rho}\frac{\partial p}{\partial z}
 \label{Eq:Axisflow-ctl}
 %\tag{\ref{Eq:2dmodel_ctl_Horizontal}}
 \end{subeqnarray}
The last one is nonhydrostatic balance in vertical direction. The
solution of above system has two parts, the rigid rotation
$(V^0_r,V^0_{\theta},V^0_z)=(0,-\frac{1}{2}fr,0)$ and the
three-dimensional flow $(V^1_r,V^1_{\theta},V^1_z)$ represented by
an axisymmetric stream function $\Psi(r,z)$. A typical value of
the rigid rotation is $\frac{1}{2}fr=1 m/s$ for
$f=5\times10^{-5}s^{-1}$ at $20^{o}N$ and $r=2\times10^{4}m$,
which is relatively small than the three-dimensional flow. It
should be noted that there is no length scale for $\Psi(r,z)$ in
Eq.(\ref{Eq:Axisflow-ctl}), thus the solution of $\Psi(r,z)$ can
be uniformly stretched by simply multiplying a real constant $a$.
We tried to find the solution of the above equations by separation
of variables $\Psi(r,z)=R(r)H(z)$. One such solution can be
written as,
\begin{equation}
V_r = \frac{ R(r)}{ r} H'(z),  \,\, V_{\theta} = \mu \frac{ R(r)}{
r} H(z)-\frac{1}{2}fr, \,\, V_{z} = - \frac{ R'(r)}{r} H(z)
 \label{Eq:Axisflow-velsol-gen}
 %\tag{\ref{Eq:2dmodel_ctl_Horizontal}}
 \end{equation}
where $'$ presents first deviation and $\mu$ is a real constant.
The absolute angular momentum $M$ of the primary circulation is
defined as,
\begin{equation}
M =rV_{\theta}+\frac{1}{2}fr^2=\mu \Psi(r,z)
 \label{Eq:Axisflow-velsol-M}
 %\tag{\ref{Eq:2dmodel_ctl_Horizontal}}
 \end{equation}
So the constant $\mu$ is the ratio of the primary circulation's
absolute angular momentum to the secondary circulation flow
streamfunction. As the $M$ is a function only $\Psi$ belong, we
have the following principle (see Appendix for proof).

Theorem 1: For axisymmetrically incompressible ideal flow, there
is an intrinsic radius $r_k$, within which is the kernel of the
primary vortex. The vortex boundary $r_k$ is the frontier of the
positive and negative vorticity of the primary circulation, and
also the frontier of upward and downward flows of the secondary
circulation. The maximum velocity of the primary circulation
locates within the vortex kernel, i.e., the radius of maximum wind
(RMW in TC studies, namely $r_m$) $r_m<r_k$.

The ratio of two different parts of the primary circulation is the
Rossby number
\begin{equation}
Ro=\frac{V^1_{\theta}}{V^0_{\theta}}=\frac{2\mu R(r)}{fr^2}H(z)
\label{Eq:Axisflow-velsol-Rossby}
 \end{equation}
The radius of the inner part of the TC is typically in the range
of 10 to 50 km \cite[]{Stern2009,ChanBook2010}, and the typical
value of $f$ is $5\times10^{-5} s^{-1}$ at $20^{o}N$. So the
typical value of $Ro\gg 1$ within this regime for a TC, and the
rotation effect could be approximately ignored for the inner part.
In the outer part of the TC, especially far from the TC core, the
Rossby number $Ro\ll 1$ and the rigid rotation is concern. It is
from Eq.(\ref{Eq:Axisflow-velsol-Rossby}) that the flow might be
geostrophic if $H(z)\rightarrow 0$ at a certain depth. In present
work, both $V^0_{\theta}$ and $V^1_{\theta}$ are obtained
analytically, but only the second one is nontrivial.

For the given stream function $\Psi(r,z)$, equation
(\ref{Eq:Axisflow-ctl}a) and equation (\ref{Eq:Axisflow-ctl}b) are
satisfied automatically. The path of a fluid material element can
be obtained,
\begin{subeqnarray}
\ln r({\theta}) &=& \displaystyle \frac{1}{\mu}\frac{H'}{H}\theta \\
\Psi(r,z) &=& const.
 \label{Eq:Axisflow-velsol-path}
 %\tag{\ref{Eq:2dmodel_ctl_Horizontal}}
 \end{subeqnarray}
In $r-\theta$ plan, it is a logarithmic spiral ($H'\neq0$), except
for $H'=0$ (a circle). So we called the solution is spiral
solution. In fact, the path is right on a Bernoulli surface given
by Eq.(\ref{Eq:Axisflow-velsol-path}b), on which the
streamfunction is defined \cite[]{Batchelor1967}. As the ideal
flow satisfies the Bernoulli¡¯s theorem, the total energy $E$ is
conserved, and can only be a function of $\Psi$ alone, a simple
relation being
\begin{equation}
\frac{dE}{d\Psi}= 4k^4\rho \Psi \label{Eq:Axisflow-velsol-Energy}
 \end{equation}
where $k$ is a constant. If $k=0$, the total energy is
homoenergetic. While the total energy is proportion to the kinetic
energy for $k\neq 0$, and the pressure is also proportion to the
kinetic energy.

It is noted that there are only two constants ($\mu$ and $k$)
constraining the problem. For any given pair of $\mu$ and $k$,
there might be different solutions for different $H(z)$ (e.g.,
Eq.(\ref{Eq:Axisflow-velsol-Hecos}) in Appendix) and $R(r)$ (e.g.,
Eq.(\ref{Eq:Axisflow-velsol-R,gamma!=0}) in Appendix). With the
solution of $R(r)$ and $H(z)$ (see Appendix for detail), the
streamfunction is $\Psi(r,z)=aR(r)H(z)$. In the following section,
we simply let $a=1$ or $a=-1$ without changing the universality,
as it is mentioned above.
%
%This implies the future works to understanding the interaction
%between waves and background flow from energy transport point of
%view.

\section{Special solutions}  \label{sec:sepcialsol}

\subsection{Mono-layer solution}\label{sec:mono-layer}

\begin{figure}
  \centerline{\includegraphics[width=3cm,height=3cm]{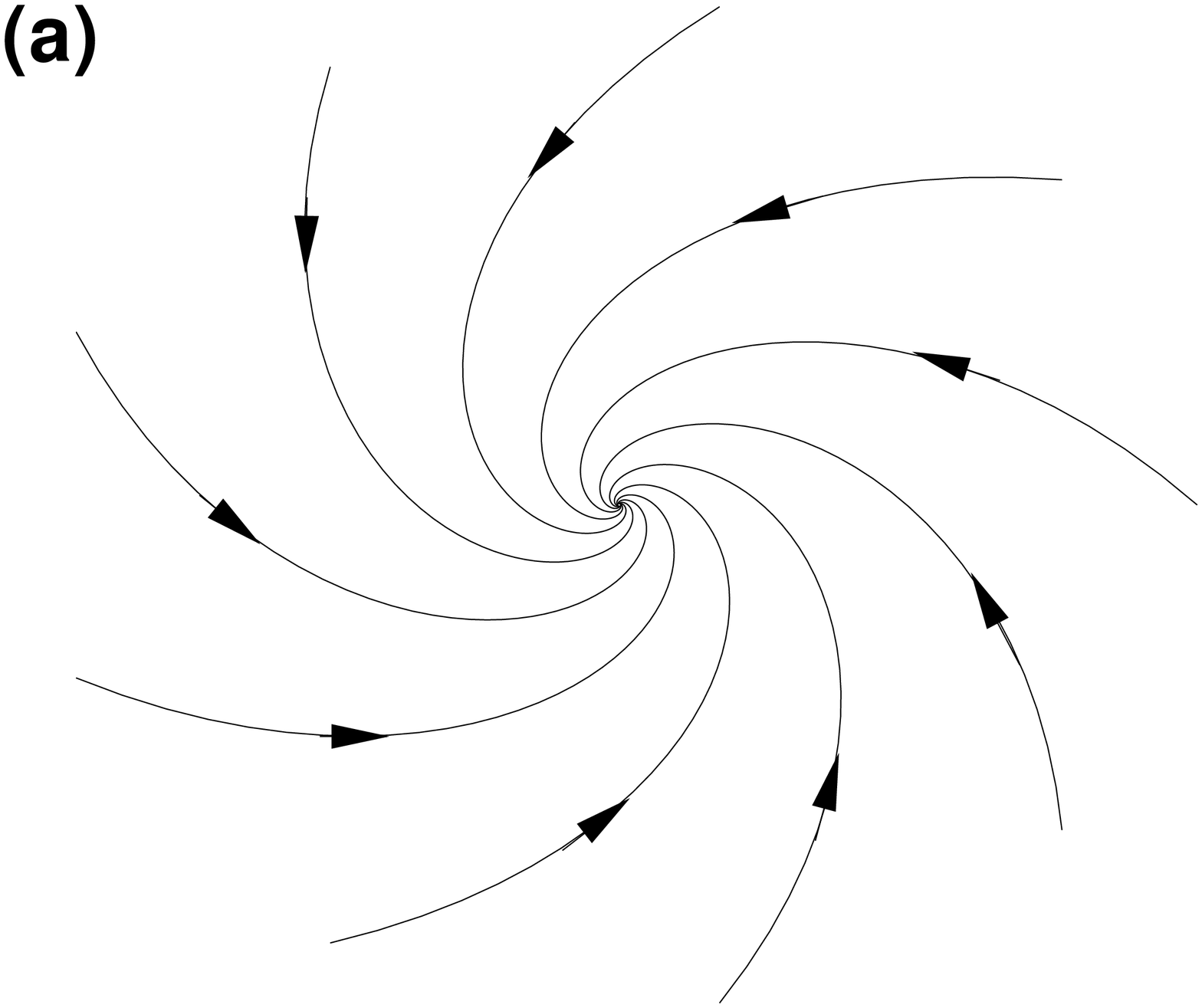}
  \includegraphics[width=3cm,height=3cm]{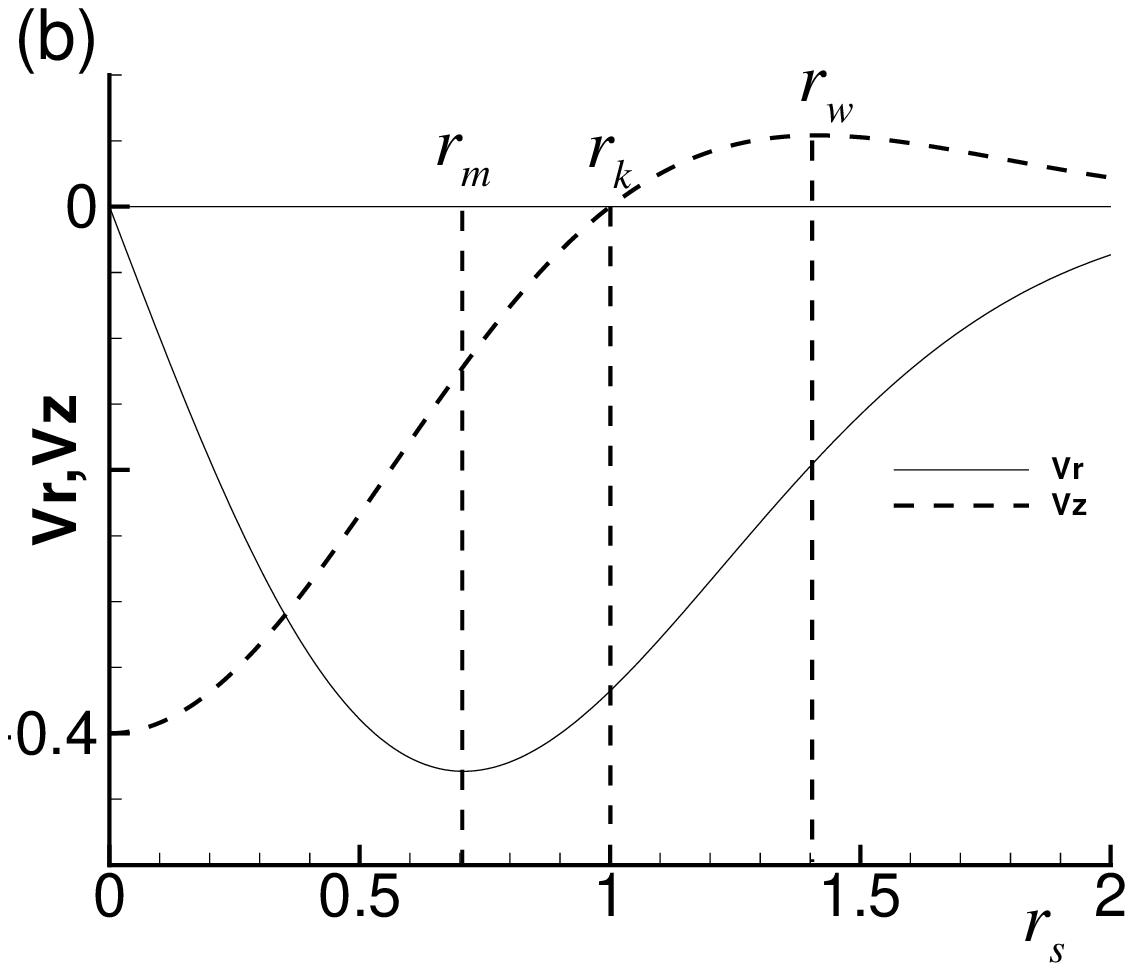}
  \includegraphics[width=3cm,height=3cm]{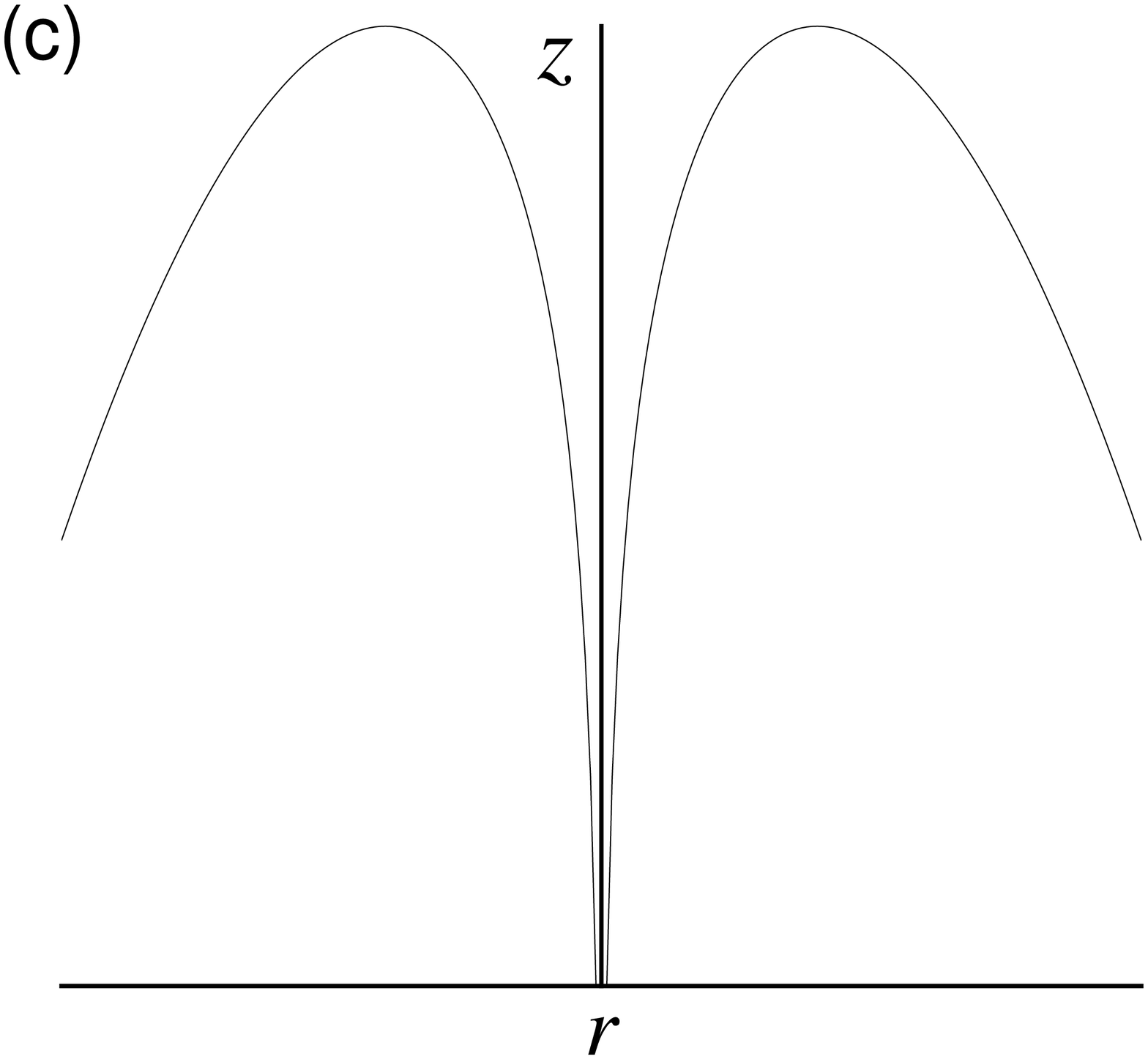}
   \includegraphics[width=3cm,height=3cm]{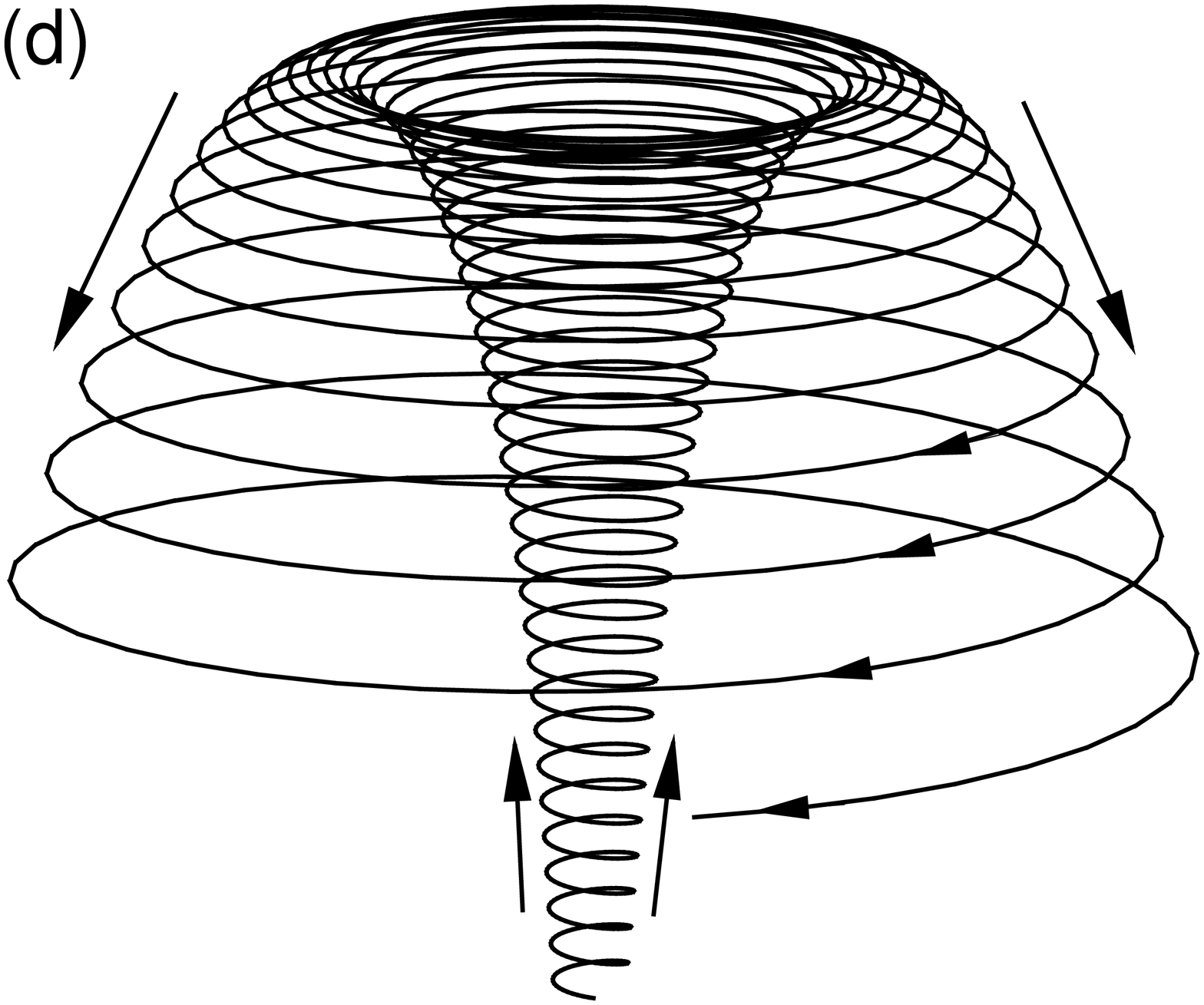}}
\caption{Paths of the fluid material elements for the TC-like
vortex. (a) $r-\theta$ plane for inflow, (b) $r-\theta$ plane for
outflow, (c) the secondary circulation in $z-r$ plane, (d) the
fluid material elements within the vortex near the kernel.}
\label{Fig:Spiral-solution-graph,gamma_i=0}
\end{figure}

In this case, the solution of $\Psi(r,z)=ar^2e^{-\lambda z-k^2
r^2}$ is finite within the whole domain, if we choose appropriate
$\lambda <0$ for $z<0$ and $\lambda
>0$ for $z>0$. If taking $\lambda >0$ for $z>0$, the radial
velocity $V_r$ does not change its direction within the domain.
Thus the secondary flow is either inflow or outflow depending on
the constant $a$ (Fig
\ref{Fig:Spiral-solution-graph,gamma_i=0}a,b). The smaller
$\lambda$ is, the higher the flow could be, as the velocity is
e-fold decline in $z$ direction. Besides, there are the standard
horizontal length scale $r_s$ and the vertical length scale $h$,
which are defined from the e-fold vertical function,
\begin{equation}
r_s=1/k,h=1/\lambda.
 \label{Eq:Axisflow-velsol-gen-rh}
 %\tag{\ref{Eq:2dmodel_ctl_Horizontal}}
 \end{equation}
The standard aspect ratio $A_s$ of the vertical length scale to
the horizontal length scale is,
\begin{equation}
A_s=h/r_s=k/\lambda.
 \label{Eq:Axisflow-velsol-gen-A}
 %\tag{\ref{Eq:2dmodel_ctl_Horizontal}}
 \end{equation}

The velocities are,
\begin{subeqnarray}
\displaystyle V_r&=& \lambda re^{-\lambda z-k^2 r^2},\\
\displaystyle V_{\theta}&=&\pm\sqrt{8k^2-\lambda^2}  re^{-\lambda z-k^2r^2}-\frac{1}{2}fr,\\
\displaystyle V_z&=& 2(1-k^2r^2)e^{-\lambda z-k^2r^2}.
 \label{Eq:Axisflow-velsol-vel,gamma!=0,ki=0}
 %\tag{\ref{Eq:2dmodel_ctl_Horizontal}}
 \end{subeqnarray}

For the primary circulation, the circular velocity of
$V^1_{\theta}$ is the same with that of the Taylor vortex for any
fixed $z$, by ignoring the rigid rotation flow $V^0_{\theta}$. The
circular velocity maximum is $V^1_{\theta m}=\sqrt{8k^2-\lambda^2}
r_m e^{-\lambda z-\frac{1}{2}}$ at $r_m^2=1/(2k^2)$, which is free
of the constants $a$, $\mu$ and $\lambda$. With the $r_m$ and
$V^1_{\theta m}$, the circular velocity could be represent as,
\begin{equation}
V^1_{\theta} =V^1_{\theta m} r_* e^{\frac{1}{2}(1-r_*^2)},
r_*=r/r_m
%\label{Eq:Axisflow-velsol-R,gamma!=0}
 %\tag{\ref{Eq:2dmodel_ctl_Horizontal}}
 \end{equation}
If the rigid rotation is taken into consideration under the
condition of $Ro\ll 1$ as $e^{-\lambda z} \ll 1$. This might
occurs at very high level for smaller $\lambda$ or very low level
for larger $\lambda$. the radius $r_m$ is slant inside
($V^1_{\theta} <0$) or outside ($V^1_{\theta} >0$) along $z$. This
can be seen from the observation data \cite[]{Stern2009}. Besides,
the primary circulation vorticity approaches to its maximum at
$r_d=2r_m$, where the secondary circulation velocity $V_z$ also
approaches its maximum value.

For the secondary circulation flow, the radial velocity reaches to
its maximum $V_{r m}=\lambda r_m e^{-\lambda z-\frac{1}{2}}$  at
$r_m^2=1/(2k^2)$. So both $V_r$ and $V^1_{\theta}$ reach their
maximums at the same radius. We could like to use the ratio of
both maximums as a scale of the primary circulation to the
secondary circulation, namely, the swirl ratio,
\begin{equation}
S=\frac{V^1_{\theta m}}{V_{r m}} =
\frac{\sqrt{8k^2-\lambda^2}}{\lambda}=\sqrt{8A_s^2-1}
\label{Eq:Axisflow-S-uniflow}
 %\tag{\ref{Eq:2dmodel_ctl_Horizontal}}
 \end{equation}
If $\lambda^2>4k^2$, then $S<1$ and the primary circulation is
weaker than the secondary circulation. However, the larger
$\lambda$ implies the larger decline in $z$ direction. So the
secondary flow is mainly restricted to near the ground of $z=0$.
On the other hand, if $\lambda^2<4k^2$, then $S>1$ the primary
circulation is stronger than the secondary circulation, the
smaller $\lambda$ is, the stronger the primary circulation is. And
the smaller $\lambda$ is (smaller decline in $z$ direction), the
higher the vortex is.

The above equation also requires $A_s>\sqrt{2}/4$, which implies
that the vertical scale must larger than a critical value for a
given horizontal length scale. On the other hand, if the vertical
scale is bounded, then the horizontal scale is also bounded.

Although the flow direction of the secondary circulation is
universal in radius, the vertical velocity $V_z$ changes its
direction at $r_k=k^{-1}=\sqrt{2}r_m$. As the line $r=r_k$
separates the upward/downward flow, we call it the radius of
vortex kernel. The paths of the fluid material elements are
illuminated in Fig \ref{Fig:Spiral-solution-graph,gamma_i=0}. In
$r-\theta$ plane (the primary circulation), the paths are cyclonic
logarithmic spirals
(Fig.\ref{Fig:Spiral-solution-graph,gamma_i=0}a) or anticyclonic
logarithmic spirals
(Fig.\ref{Fig:Spiral-solution-graph,gamma_i=0}b) with $\ln
r=-\lambda \theta/\mu $ . In $z-r$ plan (the secondary
circulation), the fluid material element moves spirally along the
surface decided by $z=(2\ln r-k^2r^2)/\lambda +const$
(Fig.\ref{Fig:Spiral-solution-graph,gamma_i=0}c). Figure
\ref{Fig:Spiral-solution-graph,gamma_i=0}d also shows the path in
the 3D space. According to the shape of the paths, we called it
the TC-like vortex. In the inner region $r<r_k$, the flow
cyclonically ascends from below to above as shown by upward arrow.
In the outer region $r>r_k$, the descends apart from the kernel.
The maximum of downward velocity locates at
$r_d=\sqrt{2}r_k=2r_m$. As $r_k$ separates the upward and downward
flow, it is the outer boundary of the eyewall, where there is the
frontier of the upward convection flow. In this TC-like vortex
solution, $r_m=\sqrt{2}r_k/2$ is always within this frontier
$r_k$. Such result is consistent with the observation mentioned in
\cite{Montgomery2011qj}.

The secondary circulation has both upward/downward flow, thus the
upward mass balances the downward mass.
\begin{equation}
\frac{2\pi r_k^2}{e} e^{-\lambda z}=-\int_{0}^{r_k} 2\pi r V_z dr
=\int_{r_k}^{\infty} 2\pi r V_z dr
 \label{Eq:Axisflow-velsol-gen-ekz,massmatch}
 %\tag{\ref{Eq:2dmodel_ctl_Horizontal}}
 \end{equation}
This implies the conservation law of mass.

In a word, this new TC-like vortex solution is very like that of
the TC structure, which we know quite limited about.

\subsection{Multi-layer solutions}\label{sec:multi-layer}

\begin{figure}
  \centerline{\includegraphics[width=3cm]{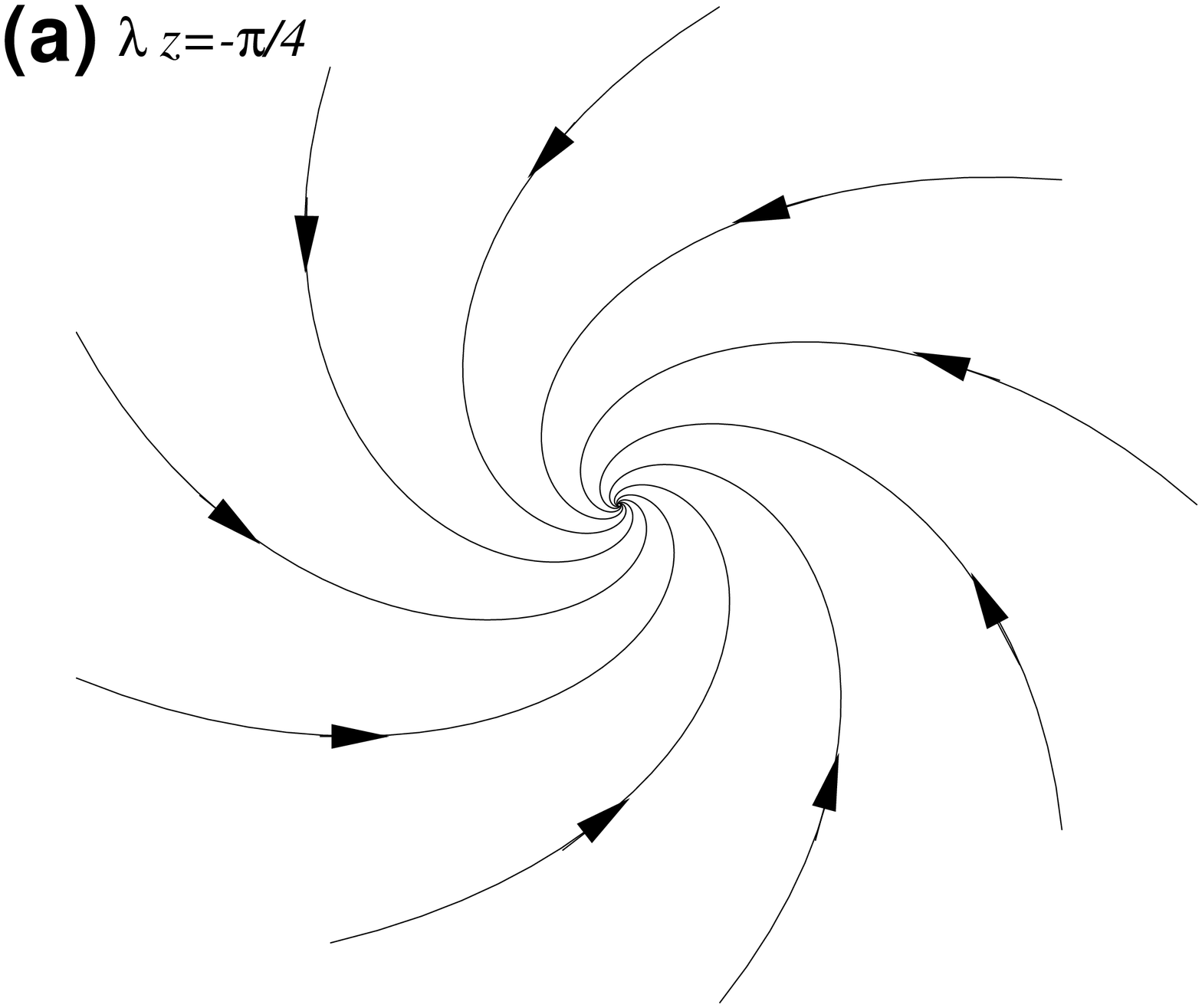}
  \includegraphics[width=3cm]{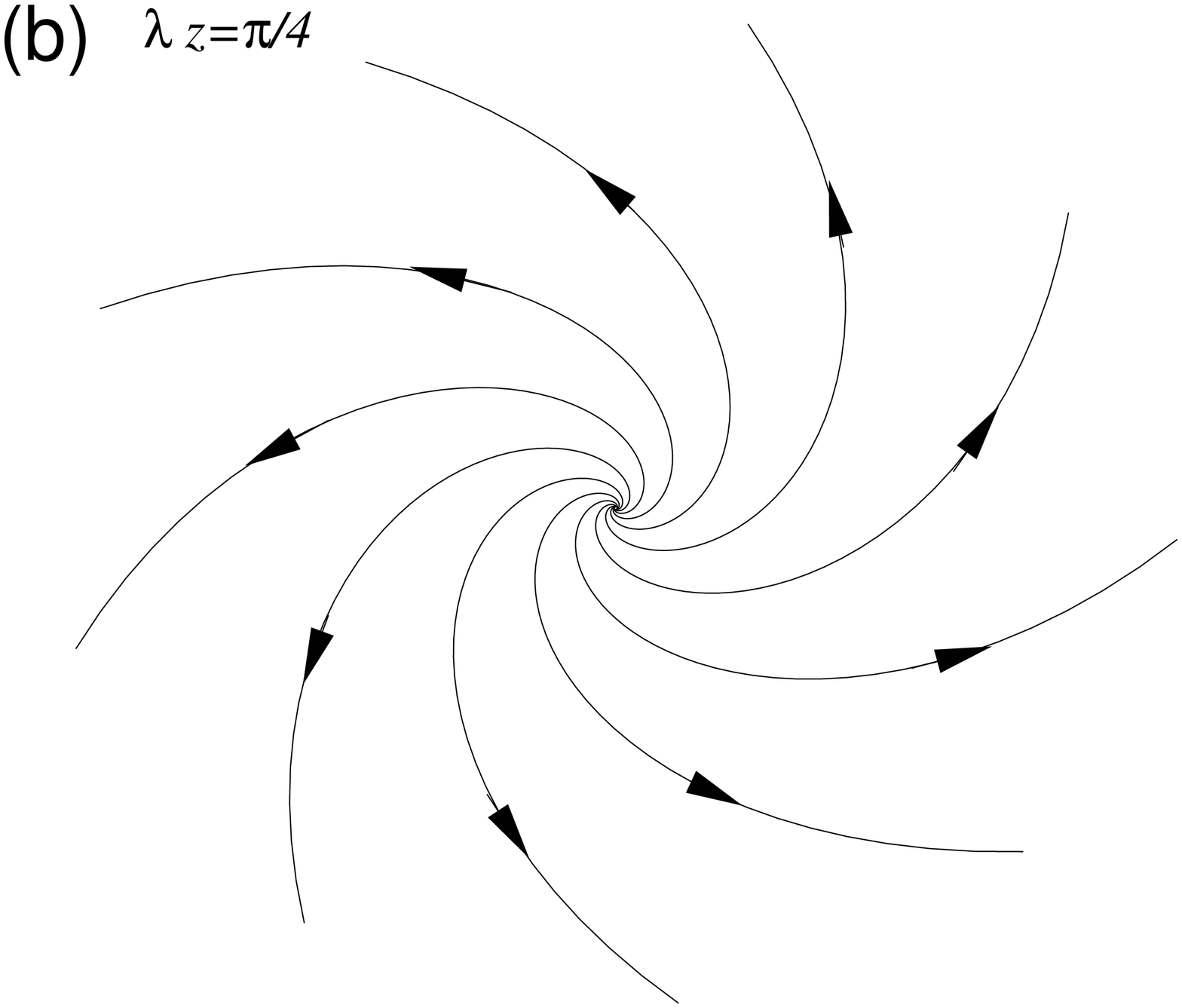}
  \includegraphics[width=3cm]{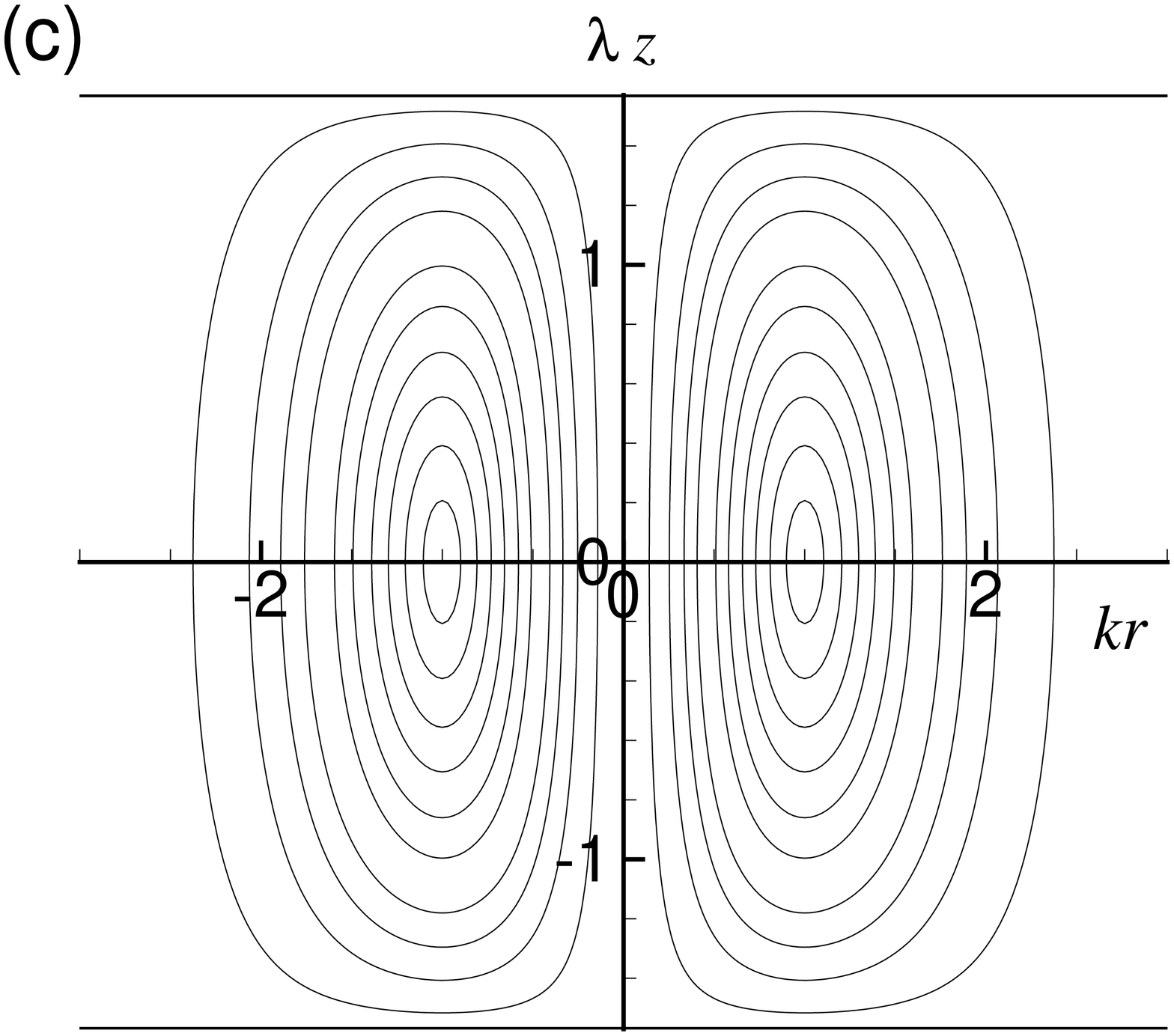}
  \includegraphics[width=3cm]{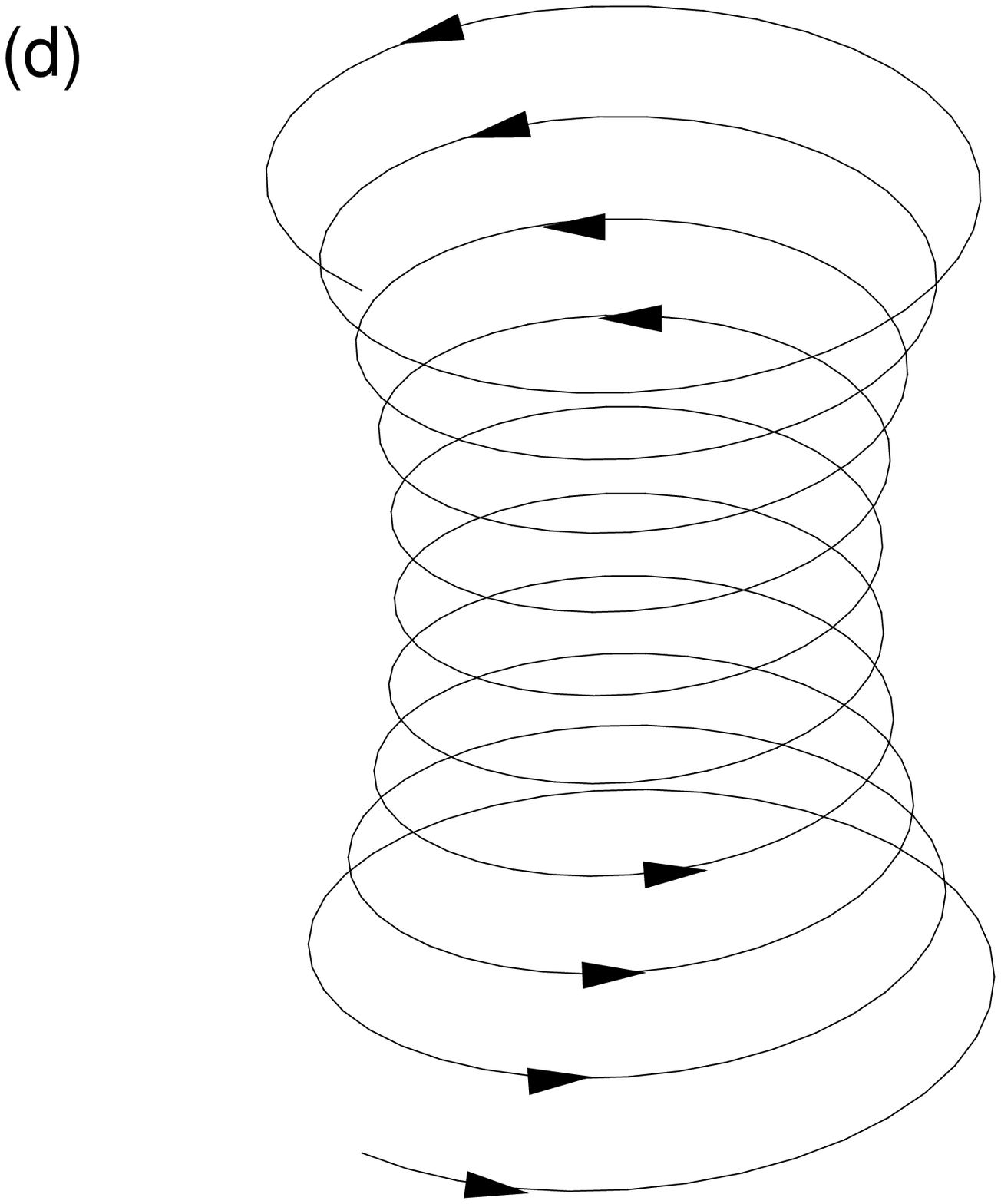} }
\caption{Paths of the fluid material elements for single-cell
vortex (vortex ring) within a layer, which is also a vortex ring.
(a) $r-\theta$ plane at $\lambda z=-\frac{\pi}{4}$, (b) $r-\theta$
plane at $\lambda z=\frac{\pi}{4}$, (c) the secondary circulation
in $z-r$ plane, (d) the spiral path of the fluid material element
within the vortex near the kernel.}
\label{Fig:Spiral-solution-graph,gamma=0}
\end{figure}

In the present solution $\Psi(r,z)=R(r)\cos(\lambda z)$  has
multiply layers by noting that the solution is periodic in
$z$-coordinate. The fluid material elements are restricted within
different vertical layers. So we call such flow as multi-planar
flow. In each layer (e.g., $-\frac{\pi}{2}\leq \lambda
z\leq\frac{\pi}{2}$), the flow has a similar behavior. It flows
inward from the far to the axis ($-\frac{\pi}{2}\leq \lambda z\leq
0$, Fig.\ref{Fig:Spiral-solution-graph,gamma=0}a), and outward
from the axis to the far ($0\leq \lambda z\leq\frac{\pi}{2}$,
Fig.\ref{Fig:Spiral-solution-graph,gamma=0}b). Thus this secondary
flow is a convection flow for $R(r)=r^2e^{-k^2r^2}$.
\begin{subeqnarray}
V_r&=&\lambda\sin(\lambda z)re^{-k^2r^2},\\
V_{\theta}&=&\pm\sqrt{8k^2+\lambda^2} \cos(\lambda z) re^{-k^2r^2}-\frac{1}{2}fr,\\
V_z&=&2\cos(\lambda z)(1-k^2r^2)e^{-k^2r^2}.
 \label{Eq:Axisflow-velsol-vel,gamma!=0,kr=0}
 %\tag{\ref{Eq:2dmodel_ctl_Horizontal}}
 \end{subeqnarray}

The primary circulation (i.e., the tangent velocity $V_{\theta}$)
has similar properties like these of the above solution. The three
intrinsic length scales are still valid for this convection flow
solution.

For the secondary circulation flow, there is a close circulation
in $z-r$ plane (Fig.\ref{Fig:Spiral-solution-graph,gamma=0}c).
This convection cell is like a vortex ring in fluid dynamics, so
it is also vortex ring flow solution. This secondary circulation
consists of radial inflow at low levels, which spirals inward
toward the vortex center, then turns up out of the boundary layer
and travels up along the vortex column, eventually expanding
outward with height. The swirl ratio becomes,
\begin{equation}
S=\frac{V^1_{\theta m}}{V_{r m}} =
\frac{\sqrt{8k^2+\lambda^2}}{\lambda \tan(\lambda
z)}=\frac{\sqrt{8A_s^2+1}}{\tan(\lambda z)}
\label{Eq:Axisflow-S-conflow}
 %\tag{\ref{Eq:2dmodel_ctl_Horizontal}}
 \end{equation}
Thus the secondary circulation is always weaker than the primary
circulation within $-\frac{\pi}{4}\leq \lambda
z\leq\frac{\pi}{4}$. The spiral path of the fluid material element
within the vortex near the kernel is depicted in
Fig.\ref{Fig:Spiral-solution-graph,gamma=0}d. As the function
$R(r)$ in Eq.(\ref{Eq:Axisflow-velsol-vel,gamma!=0,kr=0}) is the
same in Eq.(\ref{Eq:Axisflow-velsol-vel,gamma!=0,ki=0}), there are
also three intrinsic length $r_m$, $r_k$ and $r_d$, which are
right the same with these in the previous solution.

\begin{figure}
  \centerline{\includegraphics[width=6cm]{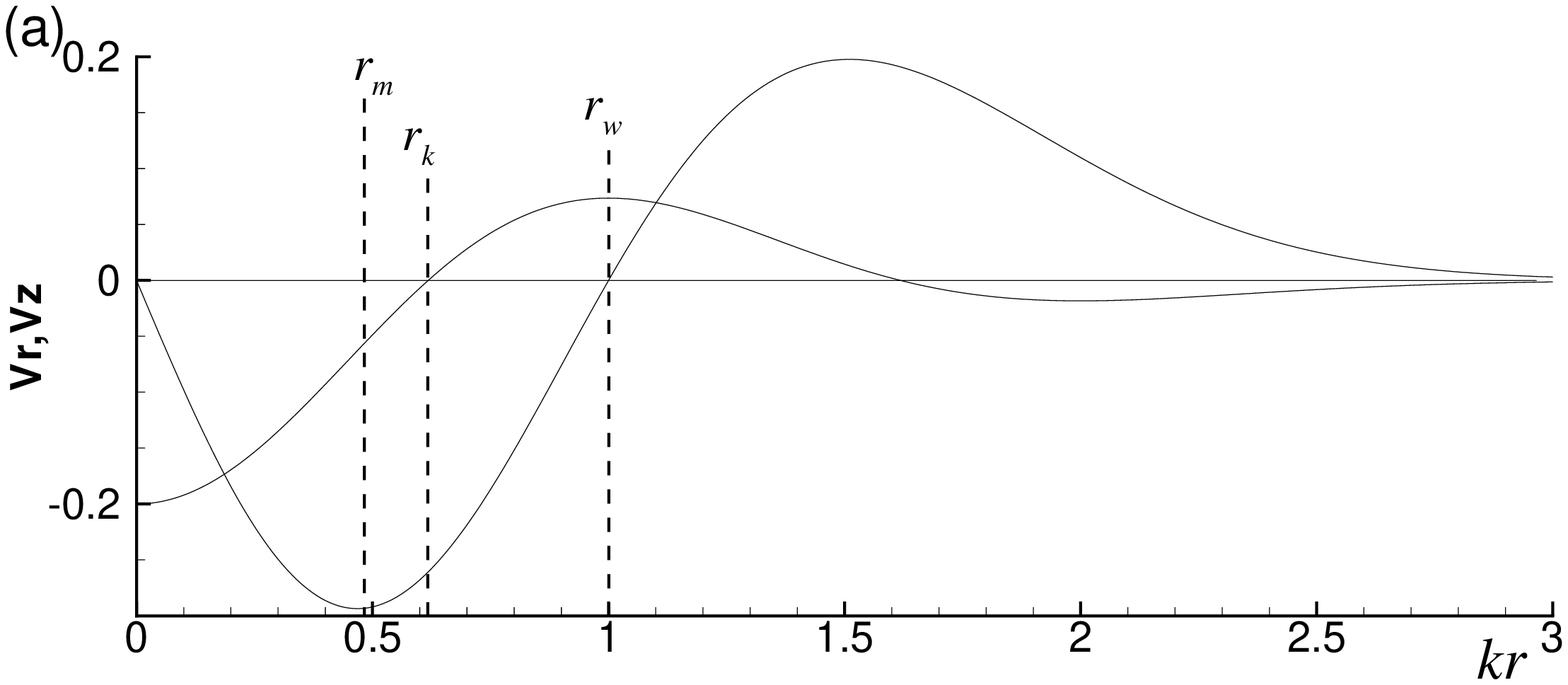}
  \includegraphics[width=6cm]{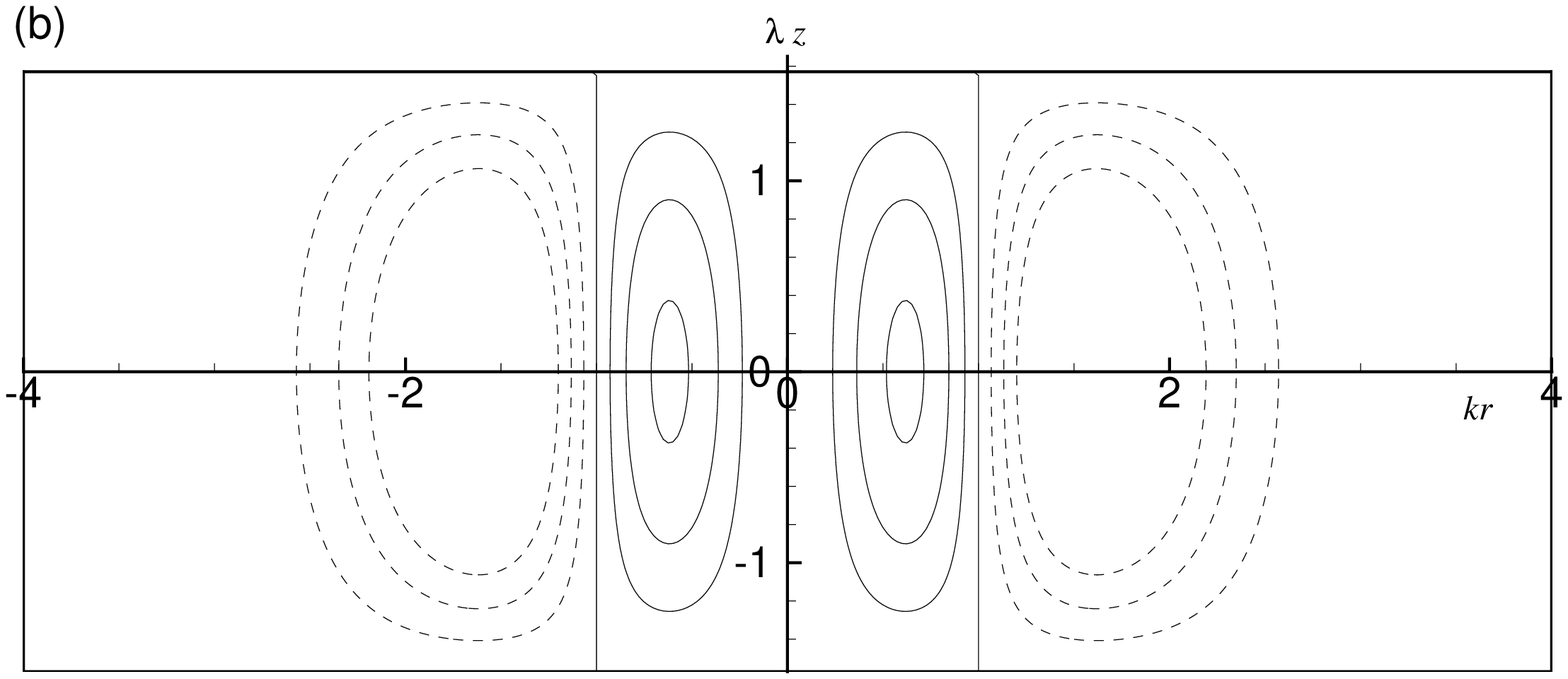} }
\caption{(a) The radius and vertical velocities for double-cell
vortex and (b) the corresponding secondary circulation in $z-r$
plane.} \label{Fig:Spiral-solution-graph,twocell}
\end{figure}
Besides, the secondary circulation could be multi-cell. In this
case, the substitution of $R$ in
Eq.(\ref{Eq:Axisflow-velsol-R,gamma!=0}) to the streamfunction
$\Psi(r,z)=P(r)R_0(r)\cos(\lambda z)$ gives the multi-cell
secondary circulation solutions, where $P_2(r)=(1-k^2r^2)$ or
$P_4(r)=(1-2k^2r^2+2/3k^4r^4)$, and etc. The flow velocity is,
\begin{subeqnarray}
V_r&=&\lambda\sin(\lambda z)\frac{P(r)R_0(r)}{r},\\
V_{\theta}&=&\mu \frac{P(r)R_0(r)}{r} \cos(\lambda z)-\frac{1}{2}fr,\\
V_z&=&\frac{[P(r)R_0(r)]'}{r}\cos(\lambda z).
 \label{Eq:Axisflow-velsol-vel,gamma!=0,k2!=0}
 %\tag{\ref{Eq:2dmodel_ctl_Horizontal}}
 \end{subeqnarray}
where $\mu=\sqrt{16k^2+ \lambda^2}$ for $P_2(r)$ and
$\mu=\sqrt{24k^2+ \lambda^2}$ for $P_4(r)$. It should be noted
that $\mu$ is different from that in the above solutions. As there
are multi-cell circulations, there are also several local maximums
for both radial velocity and tangent velocity.

For $P_2(r)=(1-k^2r^2)$, the velocities of $V_r$ and $V_z$ are
depicted in Fig.\ref{Fig:Spiral-solution-graph,twocell}a. The
two-cell solution has a close circle, and $V_r$ changes sign
across $r_0=1/k$. Thus, $r=r_0$ is a limit circle in primary
circulation: fluid outside $r_0$ will move inward to $r_0$, while
that inside $r_0$ will move outward to $r_0$
(Fig.\ref{Fig:Spiral-solution-graph,twocell}b). Therefore, near
$r_0$ there must be a strong axial flow, which in turn requires an
axial flow of opposite direction near $r=0$. Such vertical
velocity distribution is something like that of the Sullivan
vortex \cite[]{TongBGVortexBook2009,WuJZbook2006}. Comparing the
secondary circulations in
Fig.\ref{Fig:Spiral-solution-graph,twocell} with that in
Fig.\ref{Fig:Spiral-solution-graph,gamma=0}, the size of the first
circulation becomes smaller. The three intrinsic lengths are
$r_m\approx 0.468/k$, $r_k\approx 0.618/k$ and $r_d=r_0=1/k$.
Similarly, the swirl ratio is,
\begin{equation}
S=\frac{V^1_{\theta m}}{V_{r m}} =
\frac{\sqrt{16k^2+\lambda^2}}{\lambda \tan(\lambda
z)}=\frac{\sqrt{16A_s^2+1}}{\tan(\lambda z)}
\label{Eq:Axisflow-S-P2}
 %\tag{\ref{Eq:2dmodel_ctl_Horizontal}}
 \end{equation}

For $P_4(r)=(1-2k^2r^2+2k^4r^4/3)$, the velocities of $V_r$ and
$V_z$ are depicted in
Fig.\ref{Fig:Spiral-solution-graph,multicell}a. The vertical
velocity changes signs for three times. So the secondary
circulation is triple-cell
(Fig.\ref{Fig:Spiral-solution-graph,multicell}b). The intrinsic
length scales in above solution do not make sense. The three
intrinsic lengths are $r_m\approx 0.377/k$, $r_k\approx 0.496/k$
and $r_d=r_0\approx0.796/k$. Similarly, the swirl ratio is,
\begin{equation}
S=\frac{V^1_{\theta m}}{V_{r m}} =
\frac{\sqrt{24k^2+\lambda^2}}{\lambda \tan(\lambda
z)}=\frac{\sqrt{24A_s^2+1}}{\tan(\lambda z)}
\label{Eq:Axisflow-S-P4}
 %\tag{\ref{Eq:2dmodel_ctl_Horizontal}}
 \end{equation}

As calculated above, there is always $r_0=r_d$ for two-cell and
triple-cell vortex solution. This can be stated as a general
result (see Appendix for proof).

Theorem 2: For axisymmetric ideal incompressible flow, there is an
intrinsic radius $r_0$. When the radius velocity vanishes at
$r=r_0$, the vertical velocity approaches to its maximum value,
and the primary vorticity approaches to its maximum value
simultaneously, i.e., $r_0=r_d$.

\iffalse
 The radiuses of the one, two and three cell could be
approximately
\begin{subeqnarray}
r_k&=&1.256r_m+0.158r_m^2,\\
r_d&=&2.225r_m-0.226r_m^2.
 \label{Eq:Axisflow-velsol-rkrd}
 %\tag{\ref{Eq:2dmodel_ctl_Horizontal}}
 \end{subeqnarray}
\fi

\begin{figure}
  \centerline{\includegraphics[width=6cm]{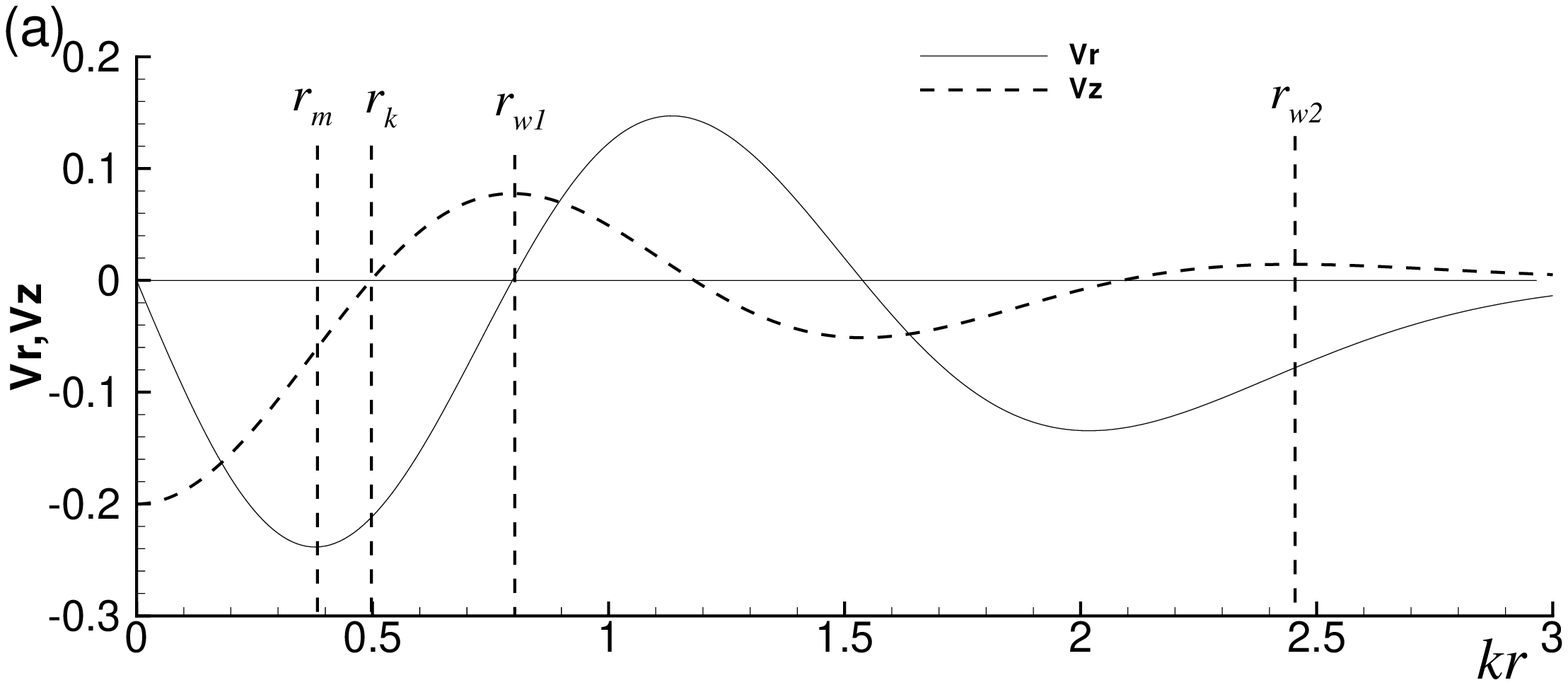}
  \includegraphics[width=6cm]{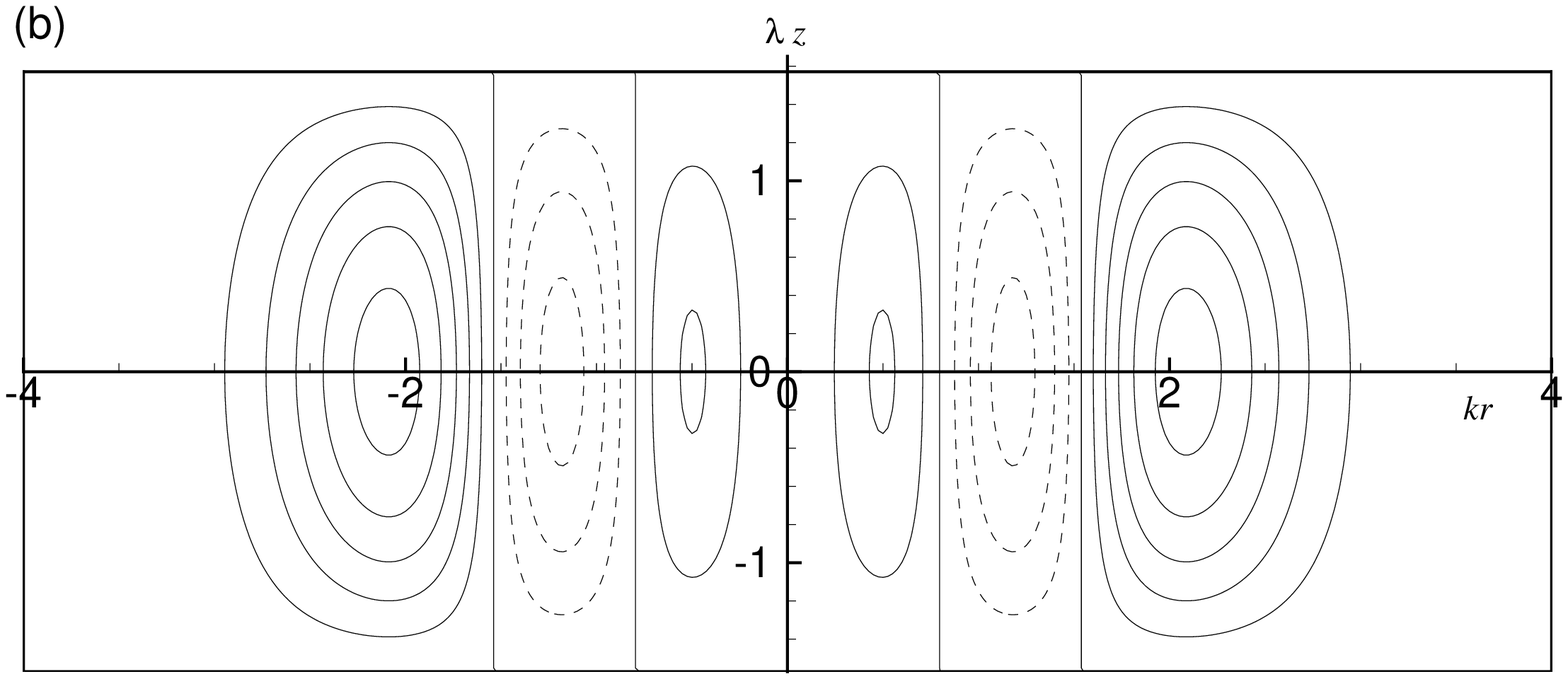} }
\caption{(a) The radius and vertical velocities for triple-cell
vortex, where $r_m\approx 0.377r_s$, $r_k\approx 0.496r_s$,
$r_{w1}=r_0\approx0.796r_s$, and $r_{w2}=r_0\approx2.45r_s$. (b)
The corresponding secondary circulation in $z-r$ plane.}
\label{Fig:Spiral-solution-graph,multicell}
\end{figure}

In a word, this kind of solution can be multi-layer and
multi-cell. There are three intrinsic length scales in the
solution. One is the radius of maximum primary circular velocity
$r_m$. The other is the radius of the vortex kernel $r_k>r_m$ for
the secondary circulation, which is also the boundary of positive
vorticity and negative vorticity for the primary circulation. The
last one is the radius of the maximum primary vorticity $r_d$, at
which the vertical flow of the secondary circulation approaches
its maximum, and across which the radius velocity changes sign.

\section{Discussion}\label{sec:discussion}

\subsection{Extensions of Solutions}
In section~\S\,\ref{sec:modelresults}, we have mentioned that the
constant $a$ could be either positive ($a=1$) or negative
($a=-1$). The different choice makes different rotation. The
secondary circulation is cyclonic in $z-r$ plane for positive $a$,
it is anticyclone for negative one. The anticyclonic one favors
the deep convection near the eyewall, where the flow
cyclonically-rotating (in $r-\theta$ plane) upward (in $z$
direction) as vortical hot tower (VHT) \cite[]{Hendricks2004}.
This may happen in the intensification of the TC. The cyclonic one
suppresses the convection, which may happen in the decline of the
TC. Both modes are intrinsic in the TC. It is the environment
condition that determine which one should be at the certain stage
of TC life.

It is also mentioned above that there are only two constants
($\mu$ and $k$) in the problem of
Eq.(\ref{Eq:Axisflow-velsol-Requaion}) and/or
Eq.(\ref{Eq:Axisflow-ctl-BHequ}). For any given pair of $\mu$ and
$k$, there might be series of different solutions for different
$\lambda$. As both equations are linear, any linear superposition
of the solutions should also satisfy the equations. So we could
use this linear superposition to obtain more useful solutions in
applications.

An alternative way to obtain new solution is that using two
different solutions combine as a new one, like the Rankine vortex.
For example, we let $a\neq0$ and $b=0$ in
Eq.(\ref{Eq:Axisflow-velsol-R,gamma=0}a) as the inner solution,
and let $a=0$ and $b\neq0$ as a outer solution. The new combine
one is a Rankine-like vortex with vertical stretch of $H(z)$.

According to the previous studies
\cite[]{Batchelor1967,Frewer2007}, the solution set is quite
large, which can also be seen from the
Eq.(\ref{Eq:Axisflow-velsol-gen}). According to
\cite{Batchelor1967}, the vertical velocity
$V_{\theta}=C(\Psi)/r$, while we simply took the velocity
components as a form  $V_{\theta}=\mu\Psi/r$ in
Eq.(\ref{Eq:Axisflow-velsol-gen}). Beyond present study, there
should be other axisymmetric solutions. For example, we can apply
the same approach to find other exact solutions by taking
$V_{\theta}=\mu/r$ or $V_{\theta}=\mu \Psi^2(r,z)/r$, etc.

\subsection{Aspect and Swirl ratios}
In the subsection \ref{sec:mono-layer}, we consider a single layer
flow with e-fold stretching in vertical axis. The aspect radio for
one-cell vortex yields to $A_s>\sqrt{2}/4$ for
$\Psi=r^2e^{-\lambda z-k^2 r^2}$ ($P_0(r)=1$ in
Eq.(\ref{Eq:Axisflow-velsol-P})\,), according to the swirl ratio
in Eq.(\ref{Eq:Axisflow-S-uniflow}). This implies that the
vertical scale must larger than a critical value for a given
horizontal length scale. On the other hand, if the vertical scale
is given, then the horizontal length scale is also bounded. For
example, if the vertical length is $h_0$, then the standard
horizontal lengthes $r_s$ yield to,
\begin{subeqnarray}
r_s&\leq&2\sqrt{2}h_0, for \, P_0\\
r_s&\leq&4h_0, for \, P_1\\
r_s&\leq&4\sqrt{2}h_0, for \, P_2
 \label{Eq:Axisflow-velsol-rs-mono}
 %\tag{\ref{Eq:2dmodel_ctl_Horizontal}}
 \end{subeqnarray}
From above relations, a bigger vortex should contain more
secondary cells. However, a bigger vortex (smaller $A_s$) also
implies a weaker swirling flow (smaller $S$ in
Eq.(\ref{Eq:Axisflow-S-conflow})\,). In contrast to that, there is
no constrain of the aspect ratio for the multi-layer flow in the
subsection \ref{sec:multi-layer}. Thus the vortex could be thin
and flat, like a thin-film.

In both cases, the swirl ratios are approximately positive proper
to the aspect radios. If the convection layer is higher, then the
swirl is weaker. On the other hand, if the vertical velocity is
strong, then this secondary convection layer must very thin
comparing to the horizontal scale.

\subsection{Possible Applications}

As mentioned above, a stretch-free inviscid vortex
(two-dimensional axisymmetric columnar vortex) can have arbitrary
radial dependence for $H(z)=1$ \cite[]{WuJZbook2006}. However,
many well-known vortex solutions are always similar to either
Eq.(\ref{Eq:Axisflow-velsol-R,gamma=0}a) or
Eq.(\ref{Eq:Axisflow-velsol-R,gamma=0}b). It is from this study
that these two-dimensional axisymmetric columnar vortices are also
the three-dimensional axisymmetric columnar vortex solutions. So
we can find either the Rankine or Taylor vortex for any fixed
layer $z=const$.

As the exact solutions are finite within the whole region, they
can be applied to study the radial structure of the tornados, the
TCs and the mesoscale eddies in the geophysical flows. As
mentioned above, there are 3 independent parameters in the
solution. To determinate them, we need the measurable qualities,
such as the radius of circular velocity maximum $r_m$, the maximum
circular velocity $V_{\theta m}$, the spiral angle defined by
$\lambda/\mu$, the angular momentum $m$, etc. According to
Eq.(\ref{Eq:Axisflow-velsol-gen}), the angular momentum
$m=rV_{\theta}=-\mu R(r)H(z)$, thus the solutions of $R(r)$
(Eq.(\ref{Eq:Axisflow-velsol-R,gamma!=0}) and
Eq.(\ref{Eq:Axisflow-velsol-R,gamma=0})) can be applied to discuss
the angular momentum of the vortex. For the tornado and TC
observations, such solutions can be used to fit the real velocity
distribution along the radial coordinate. This may also be useful
to classify the tornados and the TCs according the flow structures
provided by above solutions.

It is from Eq.(\ref{Eq:Axisflow-S-uniflow}) that the vertical flow
could be either weak (in whole height) or strong (only in
boundaries) for universal inflow solution. While from
Eq.(\ref{Eq:Axisflow-S-conflow}), the vertical flow could be
either weak (in the middle height of the region) or strong (only
near the boundaries). It seems that the relative strong radius
velocity could be easily found near boundaries.

An alternative application is the bogus TC in numerical
simulations. These new solutions have vertical velocity, which
could enhance the boundary inflow in a three-dimensionally
nonhydrostatic model.

\section{Conclusion}\label{sec:conclusion}
This investigation solve the three-dimensional ideal model to
obtain the TC-like vortex solution with a secondary circulation.
We prove two Theorems for axisymmetrically incompressible ideal
flow. First, there is an intrinsic radius $r_k$, within which is
the kernel of the primary vortex. The vortex boundary $r_k$ is the
frontier of the positive and negative vorticity of the primary
circulation, and also the frontier of upward and downward flows of
the secondary circulation. The maximum velocity of the primary
circulation locates within the vortex kernel, i.e., the radius of
maximum wind (RMW in TC studies, namely $r_m$) $r_m<r_k$. Second,
there is an intrinsic radius $r_0$. When the radius velocity
vanishes at $r=r_0$, the vertical velocity approaches to its
maximum value, and the maximum primary vorticity approaches to its
maximum value simultaneously, i.e., $r_0=r_d$. It seems that the
relative stronger radius velocity could be easily found near
boundaries.

The explicit TC-like vortex and multi-layer vortex solutions,
which are new in this investigation, might be used to describe the
3D structure of the tropical cyclones, tornados and mesoscale
eddies in the geophysical flows. The solutions can also be applied
as bogus TCs in numerical simulations.

The method used in present work is very straightforward, and it
could also be applied to other complex flows, and even for the
non-steady viscous flows like \cite{Kieu2009} tried.

The author thanks Prof. Wang W. at OUC, who discussed lots of
vortex dynamic problems with the author and encouraged the author
to finish this work. The author also thanks Dr. Wang Bo-fu and Dr.
Wan Zhen-hua for their help to check the formula, Prof. Huang
Rui-Xin at WHOI for useful comments. Prof. Yin X-Y at USTC is also
acknowledged, who led the author to this field. This work is
supported by the National Basic Research Program of China (No.
2012CB417402 and 2013CB430303), the National Foundation of Natural
Science (Nos. 41376017), and the Open Fund of State Key Laboratory
of Satellite Ocean Environment Dynamics (No. SOED1209).

\appendix
\section{Proofs}

We first prove the Theorem 1 from Eq.(\ref{Eq:Axisflow-ctl}) with
$f=0$ without loss of universality. According to
Eq.(\ref{Eq:Axisflow-ctl}a), the vertical velocity of the
secondary circulation is $V_z=-\frac{1}{r}\frac{\partial
\Psi}{\partial r}$. Integrating Eq.(\ref{Eq:Axisflow-ctl}b) with
$f=0$ gives a first integral for the momentum $M$. As $M(\Psi)$ is
a function only $\Psi$ belong, the vorticity of the primary
circulation is,
\begin{equation}
\omega=\frac{1}{r}\frac{\partial M}{\partial
r}=\frac{M'}{r}\frac{\partial \Psi}{\partial r}=-M'V_z.
 \label{Eq:Axisflow-omega}
 %\tag{\ref{Eq:2dmodel_ctl_Horizontal}}
 \end{equation}
When the vertical velocity $V_z=0$ (i.e. $\frac{\partial
\Psi}{\partial r}=0$) at a certain radius $r_k$, the vorticity
$\omega=0$ simultaneously. This is the proof of the first part of
the Theorem 1. According to the integration, the azimuth velocity
is $V_{\theta}=\frac{1}{r}M(\Psi)$. So the first deviation of
$V_{\theta}$ to $r$ is,
\begin{equation}
V'_{\theta}=\frac{\partial}{\partial
r}(\frac{M}{r})=\frac{M'}{r}\frac{\partial \Psi}{\partial
r}-\frac{1}{r}\frac{M}{r}=\omega-\frac{V_{\theta}}{r}.
 \label{Eq:Axisflow-vthetap}
 %\tag{\ref{Eq:2dmodel_ctl_Horizontal}}
 \end{equation}
At $r=r_k$, it becomes $V'_{\theta}(r_k)=-V_{\theta}(r_k)/r_k$ for
$\omega=0$. The value of $V_{\theta}(r_k)$ could be calculated by
integrating
\begin{equation}
V_{\theta}(r_k)=\int_0^{r_k} V'_{\theta}(r)dr=r_kV'_{\theta}(r_s)
 \label{Eq:Axisflow-veltheta}
 %\tag{\ref{Eq:2dmodel_ctl_Horizontal}}
 \end{equation}
where $0<r_s<r_k$. Substituting the above result to
$V'_{\theta}(r_k)$, it yields
$V'_{\theta}(r_k)=-V'_{\theta}(r_s)$. So there must be a radius
$r_m$ within $[r_s,r_k]$, where $V'_{\theta}(r_m)=0$. So we proved
the second part of the Theorem 1.

Then we solve the system of Eq.(\ref{Eq:Axisflow-ctl}).
Substituting Eq.(\ref{Eq:Axisflow-velsol-gen}) into Equation
(\ref{Eq:Axisflow-ctl}c) and equation (\ref{Eq:Axisflow-ctl}d)
becomes,
\begin{subeqnarray}
(\frac{R}{r})(\frac{R}{r})'H'^2 - (\frac{R}{r})(\frac{R'}{r})HH''-\frac{\mu^2 }{r^3}R^2H^2 =-\frac{1}{\rho}\frac{\partial p}{\partial r}  \\
(\frac{R'}{r})^2 HH' - (\frac{R}{r})(\frac{R'}{r})' HH' =
-\frac{1}{\rho}\frac{\partial p}{\partial z}
 \label{Eq:Axisflow-velsol-pressure}
 %\tag{\ref{Eq:2dmodel_ctl_Horizontal}}
 \end{subeqnarray}
Hence, the pressure can be solved from
Eq.(\ref{Eq:Axisflow-velsol-Energy}),
\begin{equation}
p(r,z)=E-\frac{\rho}{2}V^2=\frac{\rho}{2}(4k^4R^2H^2-\frac{\mu^2R^2}{r^2}H^2-\frac{R^2}{r^2}H'^2-\frac{R'^2}{r^2}H^2)
 \label{Eq:Axisflow-presol}
 %\tag{\ref{Eq:2dmodel_ctl_Horizontal}}
 \end{equation}
Substituting Eq.(\ref{Eq:Axisflow-presol}) into
Eq.(\ref{Eq:Axisflow-velsol-pressure}a) and
Eq.(\ref{Eq:Axisflow-velsol-pressure}b), both yield
\begin{equation}
(R''-\frac{R'}{r}+\mu^2R-4k^4r^2R)H+RH''=0
 \label{Eq:Axisflow-velsol-cross}
 %\tag{\ref{Eq:2dmodel_ctl_Horizontal}}
 \end{equation}
The above equation is a special case of Bragg-Hawthorne equation
\cite[]{Batchelor1967,Frewer2007},
\begin{equation}
\Psi_{zz}+\Psi_{rr}-\frac{1}{r}\Psi_r=r^2G(\Psi)+F(\Psi)
 \label{Eq:Axisflow-ctl-BHequ}
 %\tag{\ref{Eq:2dmodel_ctl_Horizontal}}
 \end{equation}
Here both $F=-\mu^2\Psi$ and $G=4k^4\Psi$ are linear functions of
$\Psi$, although $F$ and $G$ could be nonlinear in general
\cite[]{Batchelor1967,Frewer2007}.

Recall that $R$ and $H$ are independent, we can solve the above
equation. There are three kind of functions for $H(z)$ in
Eq.(\ref{Eq:Axisflow-velsol-cross}). The first trivial one is
$H=1$, any differentiable function $R(r)$ would be the solution,
e.g. the Rankine vortex \cite[]{Mallen2005,SunL2011arx}. The
second one is $H(z)=z$, one can obtains the Sullivan vortex
\cite[]{WuJZbook2006} and other vortex solutions
\cite[]{SunL2011arx}. Third, it also yields two kind of
non-trivial solutions for $H(z)$,
\begin{equation}
H(z)=e^{-\lambda (z-z_0)}\,\, or, \,\,  H(z)=\cos(\lambda
(z-z_0)). \label{Eq:Axisflow-velsol-Hecos}
 %\tag{\ref{Eq:2dmodel_ctl_Horizontal}}
 \end{equation}
where $z_0$ is a constant parameter. We simply take $z_0=0$ in the
following investigation. Substitution
Eq.(\ref{Eq:Axisflow-velsol-Hecos}) into
Eq.(\ref{Eq:Axisflow-velsol-cross}) to eliminate the function for
$z$ (plus for $H(z)=e^{\lambda z}$ and minus for $
H(z)=\cos(\lambda z)$, respectively), it yields to
\begin{equation}
R''-\frac{R'}{r}+(\mu^2\pm \lambda^2)R-4k^4r^2R=0
 \label{Eq:Axisflow-velsol-Requaion}
 %\tag{\ref{Eq:2dmodel_ctl_Horizontal}}
 \end{equation}
It is obvious that $\lambda=0$ reduces to the second case of
$H(z)=z$. Before solving the equation, we prove the Theorem 2. It
is from Eq.(\ref{Eq:Axisflow-velsol-cross}) or the above equation
that $R=0$ implies $R''r=R'$. Applying the velocity solution in
Eq.(\ref{Eq:Axisflow-velsol-gen}), when $V_r=0$ at $r=r_0$, then
$\partial (V_z)/\partial r=0$. This implies that when the radius
velocity vanishes at $r=r_0$, the vertical velocity approaches to
its maximum value. According to Eq.(\ref{Eq:Axisflow-omega}), the
primary circulation vorticity approaches to its maximum value
simultaneously. So we proved the Theorem 2.

\section{Solution}
Then we solve the above Equation
(\ref{Eq:Axisflow-velsol-Requaion}). When $\mu^2\pm \lambda^2=0$,
Eq.(\ref{Eq:Axisflow-velsol-Requaion}) has the solutions of
\begin{subeqnarray}
R(r) &=& ar^2-b, for\,\, k^2=0,\\
R(r) &=& ae^{-k^2r^2}, for\,\, k^2\neq 0
 \label{Eq:Axisflow-velsol-R,gamma=0}
 %\tag{\ref{Eq:2dmodel_ctl_Horizontal}}
 \end{subeqnarray}
When the energy $E$ is homogeneous ($k=0$), the solution of
Eq.(\ref{Eq:Axisflow-velsol-R,gamma=0}a) is the Rankine vortex
\cite[]{WuJZbook2006}. The solution of
Eq.(\ref{Eq:Axisflow-velsol-R,gamma=0}b) is the Oseen vortex
\cite[]{WuJZbook2006}. Let $R=r^2e^{-k^2r^2}P(r)$,
Eq.(\ref{Eq:Axisflow-velsol-Requaion}) yields
\begin{equation}
r^2P''+(3r-4k^2r^3)P'+(\mu^2\pm \lambda^2-8k^2)P=0
 \label{Eq:Axisflow-velsol-Pequaion}
 %\tag{\ref{Eq:2dmodel_ctl_Horizontal}}
 \end{equation}
And there are some polynomial solutions of $P(r)$,
\begin{subeqnarray}
P_0(r) &=& 1,\,\, \mu^2\pm \lambda^2-8k^2=0\\
P_2(r) &=& 1-k^2r^2,\,\,  \mu^2\pm \lambda^2-8k^2=8k^2 \\
P_4(r) &=& 1-2k^2r^2+\frac{2}{3}k^4r^4,\,\,  \mu^2\pm
\lambda^2-8k^2=16k^2
 \label{Eq:Axisflow-velsol-P}
 %\tag{\ref{Eq:2dmodel_ctl_Horizontal}}
 \end{subeqnarray}
The solution of Eq. (\ref{Eq:Axisflow-velsol-Requaion}) for
yields,
\begin{subeqnarray}
R_0(r) &=& r^2e^{-k^2r^2} ,\,\, \mu^2\pm \lambda^2-8k^2=0\\
R_2(r) &=& (1-k^2r^2)r^2e^{-k^2r^2},\,\,  \mu^2\pm \lambda^2-8k^2=8k^2 \\
R_4(r) &=& (1-2k^2r^2+\frac{2}{3}k^4r^4)r^2e^{-k^2r^2},\,\,
\mu^2\pm \lambda^2-8k^2=16k^2
 \label{Eq:Axisflow-velsol-R,gamma!=0}
 \end{subeqnarray}

Thus, Eq.(\ref{Eq:Axisflow-velsol-Hecos}) and
Eq.(\ref{Eq:Axisflow-velsol-R,gamma!=0}) give the exact solutions
of the flow velocity. It is obvious that there are three
independent parameters (e.g.,  $k$, $\mu$, $\lambda$) in the
solutions. And other new solutions can also be obtained by
combining the different solution at different regions, like that
of the Rankine vortex.

\bibliographystyle{jfm}
%\bibliography{MSH1}

\end{document}